# Generation of even and odd high harmonics in resonant metasurfaces using single and multiple ultra-intense laser pulses

Maxim R. Shcherbakov,[1,*] Haizhong Zhang,[2] Michael Tripepi,[3] Giovanni Sartorello,[1] Noah Talisa,[3] Abdallah AlShafey,[3] Zhiyuan Fan,[1] Justin Twardowski,[4] Leonid A. Krivitsky,[2] Arseniy I. Kuznetsov,[2] Enam Chowdhury,[3,4,5] Gennady Shvets[1,*]

**Affiliations:**

[1]School of Applied and Engineering Physics, Cornell University, Ithaca, NY 14853, USA.

[2]Institute of Materials Research and Engineering, A*STAR (Agency for Science, Technology and Research), 138634, Singapore.

[3]Department of Physics, The Ohio State University, Columbus, OH 43210, USA.

[4]Department of Material Science and Engineering, The Ohio State University, Columbus, OH 43210, USA.

[5]Department of Electrical and Computer Engineering, The Ohio State University, Columbus, OH 43210, USA.

*Correspondence to: mrs356@cornell.edu, gshvets@cornell.edu

**High harmonic generation (HHG)[1,2] opens a window on the fundamental science of strong-field light-mater interaction and serves as a key building block for attosecond optics and metrology[3,4]. Resonantly enhanced HHG from hot spots in nanostructures is an attractive route to overcoming the well-known limitations of gases and bulk solids[5–10]. We demonstrate a nanoscale platform for highly efficient HHG driven by strong mid-infrared laser pulses: an ultra-thin resonant gallium phosphide (GaP) metasurface. The wide bandgap and the lack of inversion symmetry of the GaP crystal enable the generation of even and odd harmonics covering a wide range of photon energies between 1.3 and 3 eV with minimal reabsorption. The resonantly enhanced conversion efficiency facilitates single-shot measurements that avoid material damage and pave the way to controllable transition between perturbative and non-perturbative regimes of light-matter interactions at the nanoscale.**

Traditionally, HHG has been observed in gases subjected to tunneling ionization by ultra-strong laser fields exceeding those that bind electrons to nuclei[11,12]. High ionization thresholds, inversion symmetry, and infrastructure requirements imposed by gas chambers present challenges to the development of small-footprint low-power sources integrable in existing optoelectronic platforms for efficient and broadband HHG. Solid-state materials represent an attractive alternative for tabletop HHG sources[13,14]. However, conventional approaches to HHG utilizing bulk crystals fail to simultaneously achieve high conversion efficiencies and broad spectral bandwidth owing to significant harmonics reabsorption and phase mismatch. More recently, designer nanostructures[6] have attracted considerable attention because they can potentially alleviate these problems due to locally enhanced optical "hot spot" fields through a variety of mechanisms: operation in the epsilon-near-zero regime, as in CdO,[10] high-quality-factor collective modes, which were demonstrated in Si metasurfaces[7], or plasmonic field enhancement[5]. However, several challenges to achieving highly-efficient HHG in the strong-field regime assisted by spectrally-selective metasurfaces may be identified. First, narrow- and moderate-bandgap semiconductors, with bandgap energies $\Delta_g$ that are not much larger than the laser photon energy $\hbar\omega$, are damaged at moderate laser fluences due to multi-photon absorption followed by rapid free-carrier generation[15,16]. Moreover, the overabundance of free carriers can drastically reduce the quality ($Q$) factor of a resonant metasurface[17], thereby defeating its key purpose: the creation of resonantly-driven optical hot spots. Second, harmonics absorption by opaque materials reduces the HHG-emitting volume and dramatically decreases the HHG efficiency[18]. Finally, only a subset of harmonics (odd) can be produced by centrosymmetric materials. Currently, non-centrosymmetric materials enabling even-order harmonics[9,19] have not been utilized for nanostructure-based HHG: to date, high ($N \geq 4$) harmonics have only been reported from nanostructures biased by an external dc field[8] or 2D semiconductors[20].

Therefore, it is desirable to develop a photonic platform and an optical system providing both the access to non-perturbative physics (defined by a strong perturbation by a laser pulse of the electron/hole motion in their respective conduction/valence bands), as well as the ability to use HHG as a probe of the microscopic processes inside a crystal[20–23]. Such combination of a photonic platform and optical system must meet the following conditions: (a) the electronic bandgap of the constitutive material should be sufficiently large, so that multiple harmonic orders can be utilized; (b) the optical system should enable single-shot measurements that do not suffer from the inherent limitations of multi-pulse averaging, such as long-term damage[24–26] and measurement biases (e.g., produced by a single high-intensity outlier in a train of laser pulses); and (c) the photonic structure should enable the production of nanoscale regions of a strongly-driven material phase embedded inside a weakly perturbed phase, thus opening the possibility of studying nonlocal effects in condensed matter phase without confounding laser damage.

The transition to nonlinear carrier motion occurs when the momentum gained from the laser electric field over a single period exceeds the size of the Brillouin zone of a solid material. This condition is expressed as $\beta \equiv \omega_B/2\omega > 1$,[14,21,27] where $\omega_B = eEa/\hbar$ is the Bloch oscillation frequency[28,29,30], $a$ a crystalline period, $\omega$ is the laser frequency, $\hbar$ is the reduced Planck's constant, and $E$ is the hot spot optical field. Concurrently, the injection of free carriers (FCs) into the conduction zone also takes place. The latter is governed by the dimensionless Keldysh parameter[31] $\gamma = \omega\sqrt{m^*\Delta_g}/eE$, where $m^*$ is the effective electron mass. Approximately equal to the ratio of the carrier injection time to laser period, the Keldysh parameter characterizes electron tunneling across the bandgap. Therefore, highly-efficient non-perturbative (saturated) HHG requires that $\beta, \gamma^{-1} > 1$.

The key challenge addressed by our work is finding the appropriate photonic platofrm for entering this new regime without producing large numbers of FCs that can blue-shift[32] or dampen[17] the metasurface resonance. As illustrated by Figure S1 (see Supplementary Information Section 1 for the calculation of strong-field-induced FC generation), our choices of the metasurface material and laser wavelength $\lambda = 2\pi c/\omega$ are strongly constrained if we are to access the non-perturbative regime of HHG in nanostructures.

Here, we design and fabricate an ultrathin ($\approx \lambda/10$, where $\lambda = 3.95$ μm) photonic platform for enhanced HHG – a resonant metasurface – based on a transparent, high-index, wide-bandgap semiconductor: gallium phosphide (GaP)[33–35]. The combination of high refractive index ($n \approx 3$) and mid-infrared (MIR) transparency enables highly localized "hot spots" of the electromagnetic field inside GaP-based metasurfaces, akin to those made of silicon and gallium arsenide[36,37].

Large electronic band gap ($\Delta_g^{(\mathrm{dir})} = 2.78$ eV and $\Delta_g^{(\mathrm{indir})} = 2.24$ eV $\gg \hbar\omega$) of GaP drastically reduces multi-photon absorption of MIR light (see Supplementary Information Section 1) and prevents visible HHG reabsorption up to the $N = 7$ harmonic frequency $\omega_N \equiv N\omega$. Finally, the non-centrosymmetric zincblende crystal structure of GaP enables generation of even-order harmonics from the bulk[9,23].

This selection of the laser wavelength and the underlying metasurface material enabled us to produce record-breaking unsaturated conversion efficiencies into high harmonics even in the perturbative regime of moderate laser intensity $I_{\max}^{(\mathrm{MP})} \approx 80\ \mathrm{GW/cm^2}$ in the multi-pulse (MP) illumination regime. By employing single-pulse (SP) measurements, we avoid laser-induced damage and reach the non-perturbative regime of HHG for incident laser intensities as high as $I_{\max}^{(\mathrm{SP})} \approx 480\ \mathrm{GW/cm^2}$. We observe a resonance-dependent saturation of the HHG at high estimated values of normalized Bloch oscillation frequencies ($\beta \approx 2$), opening exciting new opportunities for non-perturbative light-matter interactions at the nanoscale.

The metasurfaces for enhanced HHG (Fig. 1a) were fabricated from $400$ nm thick GaP films using thin film bonding, electron beam lithography and reactive ion etching (see Supplementary Information Section 2 for details). A scanning electron image of a typical metasurface sample is shown in Fig. 1b. The metasurfaces consist of densely packed domino-shaped dielectric resonant antennas (DRAs) supporting externally excited resonant electric dipole (ED) electromagnetic modes at the nominal resonant wavelength $\lambda_{\mathrm{res}}^{(0)} = 3.95\ \mu\mathrm{m}$. These modes were experimentally identified for several metasurfaces with varying dimensions (and, correspondingly, varying resonant wavelengths $\lambda_{\mathrm{res}} = \lambda_{\mathrm{res}}^{(0)} + \delta\lambda_{\mathrm{res}}$ ) using Fourier-transform infrared (FTIR) collimated beam spectroscopy[38]. At resonances – manifested as the transmission dips in the experimental (Fig. 1d) and numerical (Fig. 1e) spectra due to the excitation of the ED modes of the DRAs – metasurfaces funnel the MIR radiation into the “hot spots” (see Fig.1c for a numerical simulation). The metasurface was nominally designed to provide moderate $\left|\frac{E_{\mathrm{loc}}}{E_{\mathrm{ext}}}\right|^2 \approx 10$ intensity enhancement of the MIR radiation $\lambda = \lambda_{\mathrm{res}}^{(0)}$. The most efficient excitation of an ED mode occurs when its spectral bandwidth matches that of the MIR pump shown in Fig. 1d in gray.

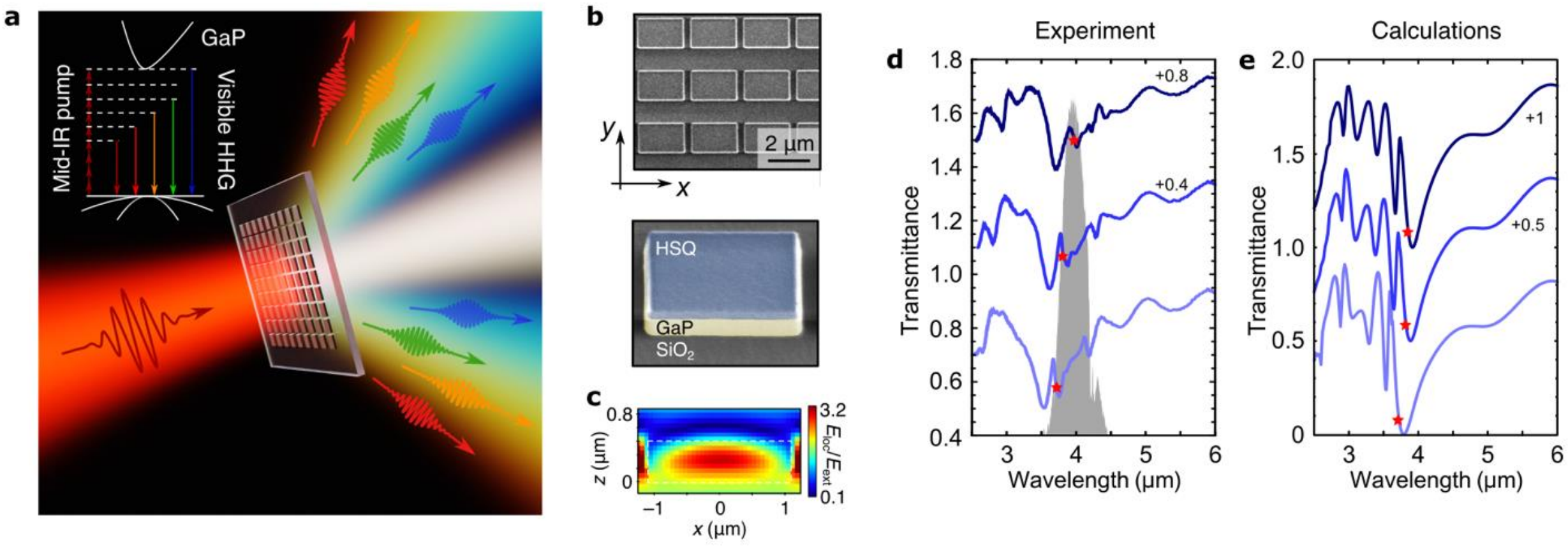

**Figure 1. GaP metasurfaces for strong-field light-matter interactions in the mid-infrared | a.** Illustration of the HHG process: resonant GaP metasurfaces show efficient even and odd high harmonic generation (up to order H9) due to the wide direct electronic bandgap, high refractive index and non-centrosymmetric lattice. The indirect band gap is not shown. **b.** Fabricated GaP metasurfaces: SEM images. **c.** Calculated local field map of the metasurface mode excited by a MIR pulse with $\lambda = \lambda_{\mathrm{res}}^{(0)}$; peak local field enhancement: $|E_{\mathrm{loc}}/E_{\mathrm{ext}}|^2 \approx 10$ at resonant wavelengths. **d.** Collimated (normal incidence) FTIR transmission spectra of three samples with varying DRA sizes: largest (upper curve) to the smallest (lower curve) size. The second and third data sets are offset for clarity by +0.4 and +0.8, respectively. **e.** COMSOL simulations of **d**. The second and third data sets are offset for clarity by +0.5 and +1.0, respectively. Red stars indicate the estimated wavelengths of the maximum local field enhancement.

Figure 2a shows a simplified sketch of the experimental setup for the detection and spectroscopy of HHG. Visible high harmonics are emitted from the metasurfaces driven by a femtosecond ($\tau_{\mathrm{MIR}} \approx 200$ fs) pulse train centered at a wavelength $\lambda = 3.95$ µm from a MIR optical parametric oscillator. The harmonics detection was performed via back focal plane (BFP) imaging or with a visible spectrometer; see Supplementary Information Sections 3-5 for details. A typical HHG spectrum, with the luminescence background subtracted, is shown in Fig. 2b. Even- and odd-order harmonics are observed in the near-infrared and visible ranges: from $\hbar\omega_4 \approx 1.2$ eV to $\hbar\omega_9 \approx 3.0$ eV (where $\omega_N = 2\pi Nc/\lambda$ is the N'th harmonic frequency). No detectible harmonic signal was observed from either unstructured GaP film of the same thickness, or the $SiO_2/Al_2O_3$ substrate. The power of the 7$^{th}$ harmonic (H7) emitted from the sample was calibrated using an external laser source of the known power and a similar wavelength; see Supplementary Information Section 7 for the calibration procedure details. The absolute conversion efficiency reaches a value of $\eta_7 \sim 2 \cdot 10^{-9}$ for H7 at $I = 80$ GW/cm$^2$, i.e. two orders of magnitude larger than the previous demonstration in a metasurface[7] and more than one order of magnitude larger than that in an epsilon-near-zero material[10].

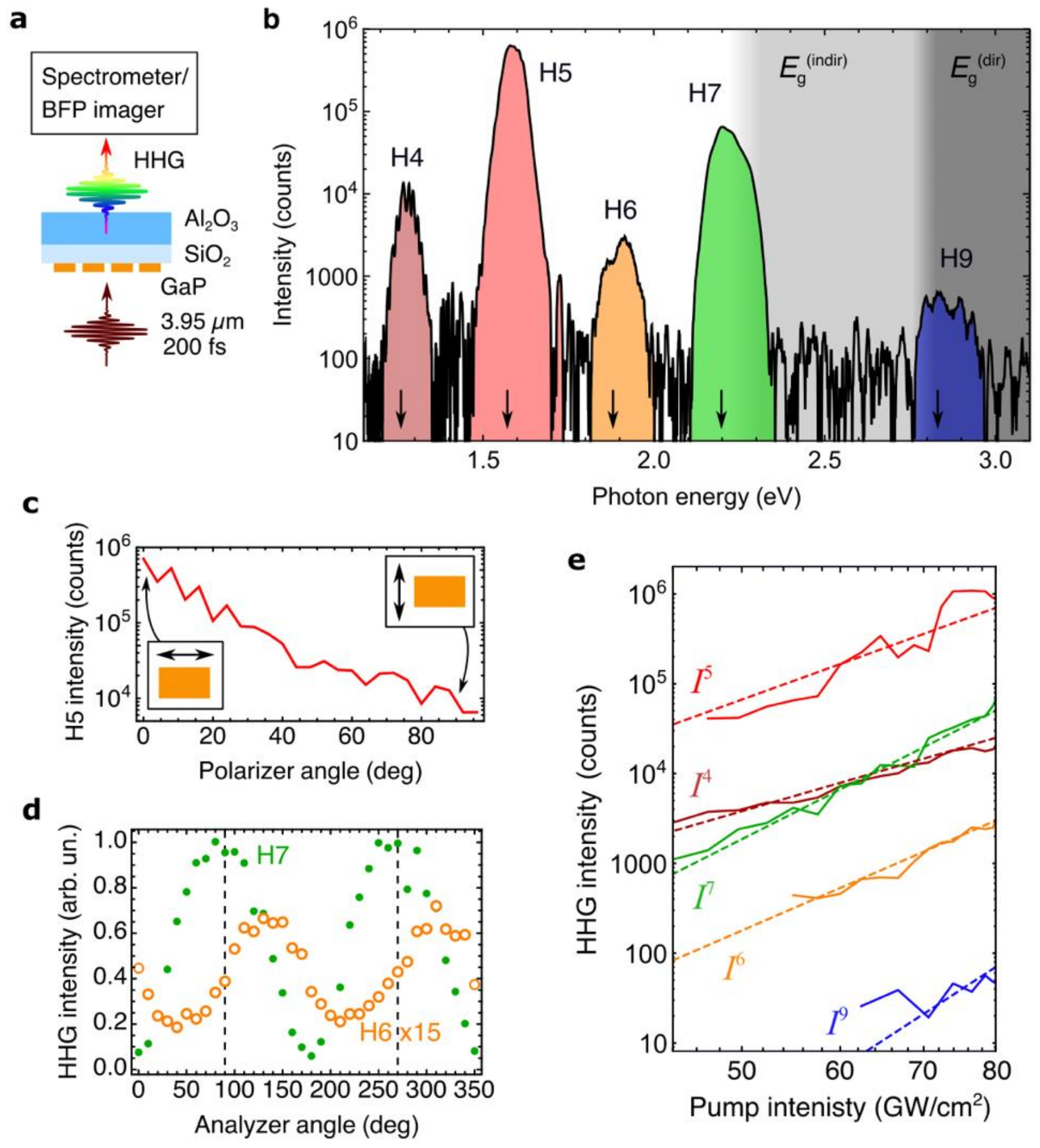

**Figure 2. High harmonic generation in the perturbative multi-pulse (MP) regime | a.** Simplified schematic of the HHG detection setup, with the detection arm represented by either a spectrometer or a back focal plane (BFP) imager. **b.** MP-HHG spectra of the resonant sample at $I_{\mathrm{MIR}} = 80\ \mathrm{GW/cm^2}$. The $N$ = 8 harmonic is not observed due to the onset of indirect interband transitions in GaP. The arrows indicate the predicted HHG wavelengths. **c.** Polarization dependence of H5 shows two orders of magnitude contrast between the resonant (horizontal) and non-resonant (vertical) MIR polarizations with $I_{\mathrm{MIR}} = 100\ \mathrm{GW/cm^2}$. **d.** Linear polarization of the odd-order (H7: green dots) and elliptic polarization of the even-order (H6: orange circles) harmonics. Dashed lines: MIR laser pulse polarization ($I_{\mathrm{MIR}} = 80\ \mathrm{GW/cm^2}$). **e.** Solid lines: HHG intensity as a function of the pump intensity for the $N$ = 4 (dark-red), $N$ = 5 (red), $N$ = 6 (orange), $N$ = 7 (green), and $N$ = 9 (blue) orders. Dashed lines: corresponding guide-to-the-eye power laws, $I^{(N)} \sim I_{\mathrm{MIR}}^{N}$.

Crucially, even-order (H4 and H6) harmonics were detected alongside the odd-order harmonics (H5, H7 and H9) because of the non-centrosymmetric (zinblende) crystal structure of GaP. Note that H8 was not detected in our experiment because of the combination of the indirect transitions at $\hbar\omega_8 = 2.28$ eV (making GaP partially opaque at H8) and the inherently lower conversion efficiency of the even-order harmonics. The relatively low efficiency of the even harmonics can be attributed to unfavorable orientation of the GaP principal crystalline axes inside the DRAs; it can be improved by about two orders of magnitude by a judicious choice of the crystal axis orientation (see Supplementary Information Section 8 and Supplementary Fig. S6).

To validate the importance of the dipole-active metasurfaces resonances, we have investigated the dependence of the H7 conversion efficiency on the polarization of the MIR pulse. The non-resonant pump polarization along the short side of the metasurface DRAs results in the efficiency reduction by two orders of magnitude compared with the resonant one as shown in Fig. 2c. This implies that optical field enhancement inside the hot spot produced by the resonant laser polarization aligned with the dipole moment of the ED mode is essential for the high efficiency of HHG observed in our experiments.

Next, we have analyzed the polarization states of the odd- and even-order harmonics. Specific examples for H7 and H6 harmonics are plotted in Fig. 2d for the (1, 0) diffraction order, as measured by BFP imaging. We observe that the odd harmonics (green dots) are co-polarized with the MIR pump (dashed lines). In contrast, the even harmonics (orange circles) are found to be elliptically polarized owing to the highly asymmetric structures of the even-order nonlinear susceptibility tensors $\chi^{(N)}_{ij\ldots k}$,[39] where the $N^{\text{th}}$-order nonlinear polarization density of the medium is given by $P_i^{(N)} = \chi^{(N)}_{ij\ldots k} E_j \ldots E_k$ (see Supplementary Information Section 8 for details). For odd values of $N$, the diagonal matrix elements of $\chi_{ij\ldots k}^{(N)}$ dominate, and the $N$th harmonic polarization is collinear with that of the MIR pump. In contrast, for even $N$, the elements of the $\chi_{ij\ldots k}^{(N)}$ tensor are predominantly off-diagonal, thereby enabling polarization changes of the even-order harmonics.

To investigate whether the HHG in the multi-pulse (moderate peak power) regime obeys the perturbative scaling laws, we have plotted in Fig. 2e the dependences of the harmonic intensity $I^{(N)}$ on the MIR intensity $I_{\mathrm{MIR}}$. The unsaturated dependences $I^{(N)} \sim I_{\mathrm{MIR}}^{N}$ are plotted as the guides for the eye. In striking difference with the previous findings of HHG in nanostructures[7,10,40], the response of the GaP metasurface does not exhibit any appreciable saturation. We conclude that the perturbative regime of harmonics generation persists up to the maximum pump intensity ($I_{\mathrm{MIR}} \approx I_{\mathrm{max}}^{(\mathrm{MP})} = 80\ \mathrm{GW/cm^2}$) used in these experiments, which is equivalent to the hot spot intensity $I_{\mathrm{hs}} \approx 0.7\ \mathrm{TW/cm^2}$ inside the metasurface. This is in agreement with our estimates of $\beta < 1$ for this range of intensities (see Table S1 of the SOM).

Because metasurfaces subjected to multi-pulse trains were visibly damaged for incident laser intensities of order $I_{\mathrm{MIR}} \approx 200\ \mathrm{GW/cm^2}$, the only non-destructive pathway to accessing the non-perturbative regime of laser-matter interaction is to resort to single-pulse (SP) experiments. Moreover, unlike MP averaging that may not provide the full picture of nonlinear processes, the SP exposures yield accurate relationships between the pulse energy, HHG signal, and the excitation site within the sample while avoiding the accumulation of MP damage[24–26]. In order to access the high-intensity regime (0.2 – 0.6 TW/cm$^2$), we replaced the focusing optics and synchronized the elements of the setup. As schematically shown in Fig.3a, the OPA triggers a mechanical shutter, which directs a single laser pulse to the sample and into the pick-off detection arm. The sample resides on a three-dimensional translation stage and is monitored by a visible-light imaging system (not shown). Each area of the sample is exposed to a single laser pulse by moving it out of the laser path by 50 μm after each shot. For each shot, the trigger starts the fast camera acquisition that records BFP images of the HHG pattern; a typical single-shot BFP image is shown in Fig. 3b.

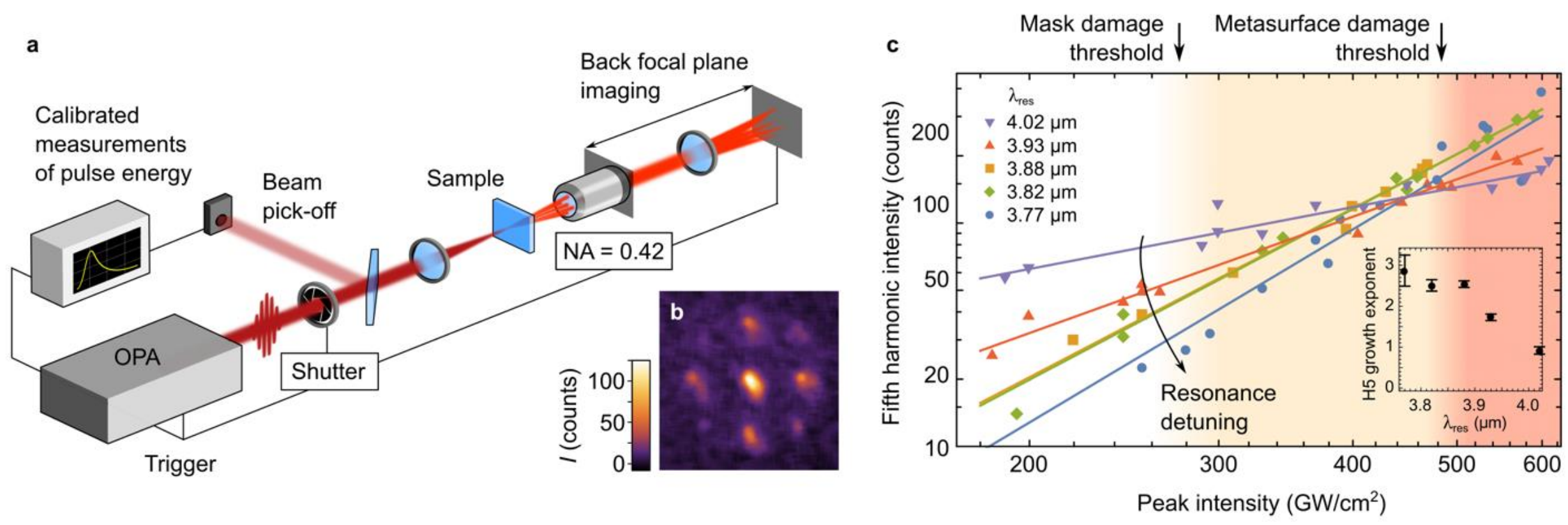

**Figure 3. Single-pulse (SP) fifth harmonic generation reveals the non-perturbative regime and high damage thresholds of resonant metasurfaces | a.** A setup for SP-HHG back focal plane (BFP) imaging. Single pulses pass through a mechanical shutter, split into the main beam (sample irradiation) and the pick-off beam (individual pulse power calibration). The diffracted harmonics are detected in the BFP configuration by triggered camera exposure. **b.** A typical BFP image of the H5 from the resonant sample at non-destructive intensities. **c.** Zeroth diffraction order intensity of the H5 as a function of MIR pump intensity for five different metasurfaces with resonances at $\lambda_{\mathrm{res}}$, from the farthest from (blue circles) to the closest to (purple triangles) the driver wavelength. Solid lines: best fits to the power law $I^{(5)} = aI^b$. Deviation from the expected $I^{(5)} \sim I^5$ indicates the saturation of nonlinear response. Inset: power exponent $b$ vs resonance wavelength $\lambda_{\mathrm{res}}$. The mask damage threshold and the metasurface damage threshold are shown for the most resonant metasurface $\lambda_{\mathrm{res}} = \lambda_{\mathrm{res}}^{(0)} = \lambda$.

As an example, the zeroth diffraction order is plotted as a function of the field intensity in Fig. 3c for five different metasurfaces, from the one with the smallest detuning between the pump and the resonance (purple triangles: $\lambda_{\mathrm{res}} = \lambda_{\mathrm{res}}^{(0)}$) to the largest detuning (blue circles). The solid lines show the best fits to the power law $I^{(5)} \sim aI^{b}$, where the exponent $b$ is expected to be equal to 5 for the perturbative H5 process. However, in contrast with moderate-intensity data in Fig. 2e, the drastic reduction of the H5 exponent ($b < 5$) signifies the onset of the non-perturbative regime. The inset of Fig. 3c shows $b(\lambda_{\text{res}})$ as a function of the detuning between the incident pulse and the resonant wavelength $\lambda_{\text{res}}$. The exponent $b$ varies monotonically between $b = 2.8$ for the least resonant metasurfaces to $b = 0.9$ for the most resonant metasurface.

The scanning electron micrographs (SEMs) of the degraded metasurfaces reveal two types of damage caused by the single pulses: the mask damage for $I_{\text{MIR}} > I_{\text{max}}^{(\text{HSQ})} \approx 280\ \text{GW/cm}^2$ (detachments of the HSQ cap from the GaP resonators) and the structure damage for $I_{\text{MIR}} > I_{\text{max}}^{(\text{GaP})} \approx 480\ \text{GW/cm}^2$ (removal of the GaP resonators from the substrate). Surprisingly, even though well-defined damage thresholds are identified by observing the metasurface degradation, no abrupt changes in HHG are experimentally observable at those threshold intensities $I_{\text{max}}^{(\text{HSQ})}$ and $I_{\text{max}}^{(\text{GaP})}$ (see Fig. 3c). The lack of any abrupt changes in the HHG dependences is attributed to the finite size of the beam: the HHG output is maintained at the beam's periphery even when the centrally positioned portion of the sample is damaged by a laser pulse. The estimated conversion efficiency of H5 in the single-pulse regime at $I = 200\ \text{GW/cm}^2$ for sample #5 (resonant case) is $\eta_{5\omega} = 1.4 \cdot 10^{-6}$, which is almost two orders of magnitude larger than that in the multi-pulse case. A comparison between various solid-state HHG sources, provided in Supplementary Table S2, shows that the GaP metasurface provides the largest specific (per unit length) conversion efficiency among all the materials provided.

One of the primary mechanisms contributing to the HHG in the non-perturbative regime is the generation of the nonlinear currents by the Bloch oscillations[29] of the FCs. The local (hot spot) field strength that does not destroy the most resonant GaP metasurface (corresponding to $I_{\mathrm{MIR}} \approx I_{\max}^{(\mathrm{GaP})}$ ) can be estimated to be $E_{\max}^{(\mathrm{hs})} \approx 0.24\, V/\text{Å}$ (assuming a factor $\times$ 10 intensity enhancement at the hot spot), bringing the value of the Bloch frequency up to $\omega_{\mathrm{B}} \approx 2 \cdot 10^{15}$ $s^{-1}$. The corresponding ratio of the Bloch frequency to the driving MIR laser frequency is $\beta = \omega_B/2\omega \approx 2.1$, thus suggesting a transition to a non-perturbative response of the underlying GaP crystal (see Table S1). The anisotropic response of the electron subsystem suggests the importance of crystal lattice orientation, whereby one can tailor the contributions from different harmonics by engineering the crystal axes with respect to the nanostructure. These effects comprise an intriguing topic for future studies.

In conclusion, we have demonstrated efficient visible high harmonics generation using mid-infrared resonances in ultra-thin gallium phosphide metasurfaces. Our approach provides record-high conversion efficiency at the nanoscale, enabled by the combination of strong hot spot enhancement of the optical field, high resilience of the underlying material to strong fields, and the low level of HHG reabsorption. Single-pulse illumination format enabled us to utilize much higher laser intensities than in the multiple-pulse format, thereby accessing the non-perturbative regime of HHG without confounding structural damage. The robustness of the metasurface to laser damage under ultra-intense illumination opens new routes to accessing strong-field regimes with tailored light fields and enables non-perturbative light-matter interactions on a chip.

**References and Notes:**

1. McPherson, A. *et al.* Studies of multiphoton production of vacuum-ultraviolet radiation in the rare gases. *J. Opt. Soc. Am. B* **4**, 595 (1987).

2. Ferray, M. *et al.* Multiple-harmonic conversion of 1064 nm radiation in rare gases. *J. Phys. B At. Mol. Opt. Phys.* **21**, (1988).

3. Paul, P. M. *et al.* Observation of a train of attosecond pulses from high harmonic generation. *Science.* **292**, 1689–1692 (2001).

4. Krausz, F. & Ivanov, M. Attosecond physics. *Rev. Mod. Phys.* **81**, 163–234 (2009).

5. Vampa, G. *et al.* Plasmon-enhanced high-harmonic generation from silicon. *Nat. Phys.* **13**, 659–662 (2017).

6. Sivis, M. *et al.* Tailored semiconductors for high-harmonic optoelectronics. *Science.* **306**, 303–306 (2017).

7. Liu, H. *et al.* Enhanced high-harmonic generation from an all-dielectric metasurface. *Nat. Phys.* **14**, 1006–1010 (2018).

8. Vampa, G. *et al.* Strong-field optoelectronics in solids. *Nat. Photonics* **12**, 465–468 (2018).

9. Liu, S. *et al.* An all-dielectric metasurface as a broadband optical frequency mixer. *Nat. Commun.* **9**, 2507 (2018).

10. Yang, Y. *et al.* High-harmonic generation from an epsilon-near-zero material. *Nat. Phys.* **15**, 1022–1026 (2019).

11. Schafer, K. J., Yang, B., Dimauro, L. F. & Kulander, K. C. Above threshold ionization beyond the high harmonic cutoff. *Phys. Rev. Lett.* **70**, 1599–1602 (1993).

12. Bartels, R. *et al.* Shaped-pulse optimization of coherent emission of high-harmonic soft x-rays. *Nature* **406**, 164–166 (2000).

13. Ghimire, S. *et al.* Observation of high-order harmonic generation in a bulk crystal. *Nat. Phys.* **7**, 138–141 (2011).

14. Ghimire, S. & Reis, D. A. High-harmonic generation from solids. *Nat. Phys.* **15**, 10–16 (2019).

15. Austin, D. R. *et al.* Femtosecond laser damage of germanium from near- to mid-IR wavelengths. *Opt. Lett.* **43**, 3702–3705 (2018).

16. Werner, K. *et al.* Single-Shot Multi-Stage Damage and Ablation of Silicon by Femtosecond Mid-infrared Laser Pulses. *Sci. Rep.* **9**, 1–13 (2019).

17. Shcherbakov, M. R. *et al.* Time-variant metasurfaces enable tunable spectral bands of negative extinction. *Optica* **6**, 1441 (2019).

18. Liu, H. *et al.* Overcoming the absorption limit in high-harmonic generation from crystals. (2019).

19. Liu, S. *et al.* Resonantly Enhanced Second-Harmonic Generation Using III--V Semiconductor All-Dielectric Metasurfaces. *Nano Lett.* **16**, 5426–5432 (2016).

20. Liu, H. *et al.* High-harmonic generation from an atomically thin semiconductor. *Nat. Phys.* **13**, 262–265 (2016).

21. Vampa, G. *et al.* Theoretical analysis of high-harmonic generation in solids. *Phys. Rev. Lett.* **113**, 073901 (2014).

22. Hafez, H. A. *et al.* Extremely efficient terahertz high-harmonic generation in graphene by hot Dirac fermions. *Nature* **561**, 507–511 (2018).

23. Hohenleutner, M. *et al.* Real-time observation of interfering crystal electrons in high-

harmonic generation. *Nature* **523**, 572–575 (2015).

24. Leitner, T. *et al.* Shot-to-shot and average absolute photon flux measurements of a femtosecond laser high-order harmonic photon source. *New J. Phys.* **13**, 093003 (2011).

25. Goh, S. J. *et al.* Single-shot fluctuations in waveguided high-harmonic generation. *Opt. Express* **23**, 24888 (2015).

26. Nisoli, M. *et al.* Effects of carrier-envelope phase differences of few-optical-cycle light pulses in single-shot high-order-harmonic spectra. *Phys. Rev. Lett.* **91**, 213905 (2003).

27. Ghimire, S. *et al.* Observation of high-order harmonic generation in a bulk crystal. *Nat. Phys.* **7**, 138–141 (2011).

28. Ghimire, S. *et al.* Strong-field and attosecond physics in solids. *J. Phys. B At. Mol. Opt. Phys.* **47**, 204030 (2014).

29. Schubert, O. *et al.* Sub-cycle control of terahertz high-harmonic generation by dynamical Bloch oscillations. *Nat. Photonics* **8**, 119–123 (2014).

30. Wu, M., Ghimire, S., Reis, D. A., Schafer, K. J. & Gaarde, M. B. High-harmonic generation from Bloch electrons in solids. *Phys. Rev. A* **91**, 043839 (2015).

31. Keldysh, L. V. Ionization in the Field of a Strong Electromagnetic Wave. *J. Exptl. Theor. Phys.* **20**, 1307–1314 (1965).

32. Shcherbakov, M. R. *et al.* Photon acceleration and tunable broadband harmonics generation in nonlinear time-dependent metasurfaces. *Nat. Commun.* **10**, 1345 (2019).

33. Rivoire, K., Lin, Z., Hatami, F., Ted Masselink, W. & Vučković, J. Second harmonic generation in gallium phosphide photonic crystal nanocavities with ultralow CW pump

power. *Opt. Express* **17**, 22609–22615 (2009).

34. Logan, A. D. *et al.* 400%/W second harmonic conversion efficiency in 14 μm-diameter gallium phosphide-on-oxide resonators. *Opt. Express* **26**, 33687 (2018).

35. Wilson, D. J. *et al.* Integrated gallium phosphide nonlinear photonics. *Nat. Photonics* **14**, 57–62 (2020).

36. Kuznetsov, A. I., Miroshnichenko, A. E., Brongersma, M. L., Kivshar, Y. S. & Lukyanchuk, B. Optically resonant dielectric nanostructures. *Science.* **354**, aag2472 (2016).

37. Koshelev, K. *et al.* Subwavelength dielectric resonators for nonlinear nanophotonics. *Science.* **367**, 288–292 (2020).

38. Wu, C. *et al.* Spectrally selective chiral silicon metasurfaces based on infrared Fano resonances. *Nat. Commun.* **5**, 3892 (2014).

39. Boyd, R. *Nonlinear Optics*. (Elsevier, 2008).

40. Yoshikawa, N., Tamaya, T. & Tanaka, K. High-harmonic generation in graphene enhanced by elliptically polarized light excitation. *Science.* **356**, 736–738 (2017).

## Methods

For the list and description of methods, see Supplementary information.

### Acknowledgements

M. R. S., Z. F., and G. Shvets acknowledge support from Office of Naval Research grant #N00014-17-1-2161. E. C., N. T., and A. S. acknowledge support from Air Force Office of Scientific Research grant (FA9550-16-1-0069) and AFOSR multidisciplinary research program of the university Research initiative (MURI) grant (FA9550-16-1-0013). MT acknowledges support from UES contract #GRT00052880. H. Z., L. A. K. and A. I. K. acknowledge support from A*STAR Quantum Technology for Engineering (QTE) program and A*STAR SERC Pharos program, Grant No. 152 73 00025 (Singapore). The authors thank Daniil Shilkin for providing the illustration in Fig. 1a.

**Author Contributions**

M. R. S., G. Shvets and A. I. K. conceived the idea. M. R. S. and Z. F. designed the sample. H. Z., L. A. K. and A. I. K. fabricated the samples. G. Sartorello measured the infrared transmittance spectra of the samples. M. R. S., M. T., G. Sartorello, N. T., A. A., J. T. and E. C. performed the nonlinear optical measurements. M. R. S. prepared the initial draft of the manuscript. A. I. K., E. C. and G. Shvets supervised the project. All authors contributed to the final version of the manuscript.

**Competing interests**

The authors declare no competing interests.

# Generation of even and odd high harmonics in resonant metasurfaces using single and multiple ultra-intense laser pulses. Supplementary information

**Authors:** Maxim R. Shcherbakov,[1,*] Haizhong Zhang,[2] Michael Tripepi,[3] Giovanni Sartorello,[1] Noah Talisa,[3] Abdallah AlShafey,[3] Zhiyuan Fan,[1] Justin Twardowski,[3] Leonid A. Krivitsky,[2] Arseniy I. Kuznetsov,[2] Enam Chowdhury,[3,4,5] Gennady Shvets[1,*]

**Affiliations:**

[1]School of Applied and Engineering Physics, Cornell University, Ithaca, NY 14853, USA.

[2]Institute of Materials Research and Engineering, A*STAR (Agency for Science, Technology and Research), 138634, Singapore.

[3]Department of Physics, The Ohio State University, Columbus, OH 43210, USA.

[4]Department of Material Science and Engineering, The Ohio State University, Columbus, OH 43210, USA.

[5]Department of Electrical and Computer Engineering, The Ohio State University, Columbus, OH 43210, USA.

*Correspondence to: mrs356@cornell.edu, gshvets@cornell.edu

## 1. Experimental parameter and ionization rates estimates

Following the estimates of the field intensity and the temporal local field enhancement factor $L \approx 3$ (defined as $E = LE_{\text{ext}}$, where $E$ is the peak value of the field within the GaP resonator, $E_{\text{ext}}$ is the external field), we can provide estimates of the peak fields within the structure, as well as the general metrics of the light-matter interactions in the metasurface. In Supplementary Table S1, three values of field intensity are given: the maximum multi-shot intensity used in Fig. 2e, minimum single-shot intensity used in Fig. 3 of the main text, and the intensity corresponding to the middle point between the damage thresholds of the mask and the resonators. The columns show the following calculated quantities: vacuum intensity $I_{\text{vac}}$, local intensity $I = L^2 I_{\text{vac}}$, electric field strength $E = \sqrt{I[\text{TW/cm}^2]}\ 0.137\ \text{V/Å}$; the Bloch oscillation frequency $\omega_{\text{B}} = |eE|a/\hbar$, where $e = 1.6 \times 10^{-19}$ C, $a = 5.44$ Å is the lattice constant of GaP, and $\hbar = 1.05 \times 10^{-34}$ J · s is the reduced Planck's constant; $\beta$ parameter, $\beta = \omega_{\text{B}}/2\omega$, where $\omega = 4.77 \cdot 10^{14}\ \text{s}^{-1}$ is the pump frequency; $\delta$ parameter, $\delta = E/E_{\text{crit}}$, where the critical field $E_{\text{crit}} = \Delta_g/ea$ and $\Delta_g = 2.78$ eV is the Γ-point gap; the Keldysh parameter $\gamma = \omega\sqrt{m^*\Delta_{\text{g}}}/eE$, where $m^* = 0.09m$ is the Γ-valley effective electron mass, $m = 9.1 \cdot 10^{-31}$ kg.

| $I_{\text{vac}}$ (TW/cm$^2$) | $I$ (TW/cm$^2$) | $E$ (V/Å) | $\omega_{\text{B}}$ ($10^{14}$ s$^{-1}$) | $\beta = \frac{\omega_B}{2\omega}$ | $\delta = \frac{E}{E_{\text{crit}}}$ | $\gamma$ |
|---|---|---|---|---|---|---|
| 0.07 | 0.6 | 0.11 | 9 | 0.9 | 0.21 | 0.52 |
| 0.2 | 1.8 | 0.18 | 15 | 1.6 | 0.36 | 0.31 |
| 0.33 | 3 | 0.24 | 20 | 2.1 | 0.46 | 0.24 |

**Supplementary Table S1 | Light-matter interaction metrics in gallium phosphide metasurfaces.** $I_{vac}$ is the MIR pump intensity in a vacuum near the focal spot ($\lambda = 3.95\ \mu m$), $I$ is the pump intensity at the metasurface's hot spot, $E$ is the MIR pump field strength at the metasurface's hot spot, $\omega_B = aeE/\hbar$ is the corresponding Bloch frequency, $E_{crit} = e\Delta_g/a, \gamma = \omega\sqrt{m^*\Delta_g}/eE$ is the Keldysh parameter. Note that for high field strengths, the effective mass approximation may not provide reliable values of $\gamma$.

To estimate the ionization rates in a material with the ionization potential $\Delta_g$ by an ac field at frequency $\omega$ and strength $E$, we apply Perelomov–Popov–Terent'ev theory[S1] which describes both the multiphoton and tunneling ionization processes. In this model, time-dependent Schrodinger equation is analytically solved in the quasi-static, single active electron approximation. We estimate the ionization rate using the following solution:

$$W \approx \frac{\Delta_g}{\hbar}\left(\frac{E\sqrt{1+\gamma^2}}{2E_0}\right)^{\frac{3}{2}} A(\omega,\gamma)e^{-\frac{2E_0}{3E}g(\gamma)},$$

where $A(\omega,\gamma) = \gamma^2/(1+\gamma^2)\sum_{n\leq\nu}\exp[-\alpha(\gamma)(n-\nu)]\,w[\sqrt{\beta(n-\nu)}]$, $\alpha = 2(\operatorname{arcsinh}\gamma - \gamma/\sqrt{1+\gamma^2})$, $\beta = 2\gamma/\sqrt{1+\gamma^2}$, $w(x) = \frac{x}{2}\int_0^1 \exp(-x^2t)\,t/\sqrt{1-t}dt$, $g = 3/2\gamma\left[(1+1/2\gamma^2)\operatorname{arcsinh}\gamma - \sqrt{1+\gamma^2}/2\gamma\right]$, and $E_0$ is the characteristic field of an electron in a state with binding energy $E_g$. Calculated rates $W$ are multiplied by the pulse duration $t = 100$ fs to estimate the ionization probability by a single pulse $P = Wt$, which is plotted in Supplementary Fig. S1 as a function of the excitation field frequency for various values of $\beta = aeE/2\hbar\omega$, where $a = 5.44$ Å is the crystal lattice constant. The regime with $\beta > 1$, whereby Bloch oscillations can meaningfully contribute to the process of HHG, is denoted by the shaded pink area. We have also estimated the ionization probability that causes more than $\rho_{\text{crit}} = 10^{19}$ cm$^{-3}$ electron-hole pairs at $P_{\text{crit}} = \rho_{\text{crit}}a^3/8 \approx 10^{-4}$, where 8 stands for the number of atoms in the unit cell, as marked by the dashed line in Supplementary Fig. S1. The plasma as dense as $\rho_{\text{crit}}$ can cause significant free-carrier-induced absorption and deteriorate both the resonance of the metasurface and its HHG output. The circle indicates the estimated experimental conditions ($\beta \approx 2, \hbar\omega/\Delta_g \approx 0.12$), which signify an important physical regime, whereby an ultra-intense laser pulse can lead to non-perturbative response without actively damaging the material or causing sub-optimal nonlinear response hindered by the generated free carriers. The family of dashed curves plotted for a narrower band gap of 1 eV show a considerably higher ionization, which may prevent efficient HHG in narrow-gap semiconductors. It is important to note that, even though the PPT model is not as accurate as the Keldysh model,[S2] because it does not take into account any details of the electronic band structure, we have verified that the two model are in approximate agreement for the case of GaP, as the Keldysh model provides us with similar ionization rates for the parameters we used in our experiment.

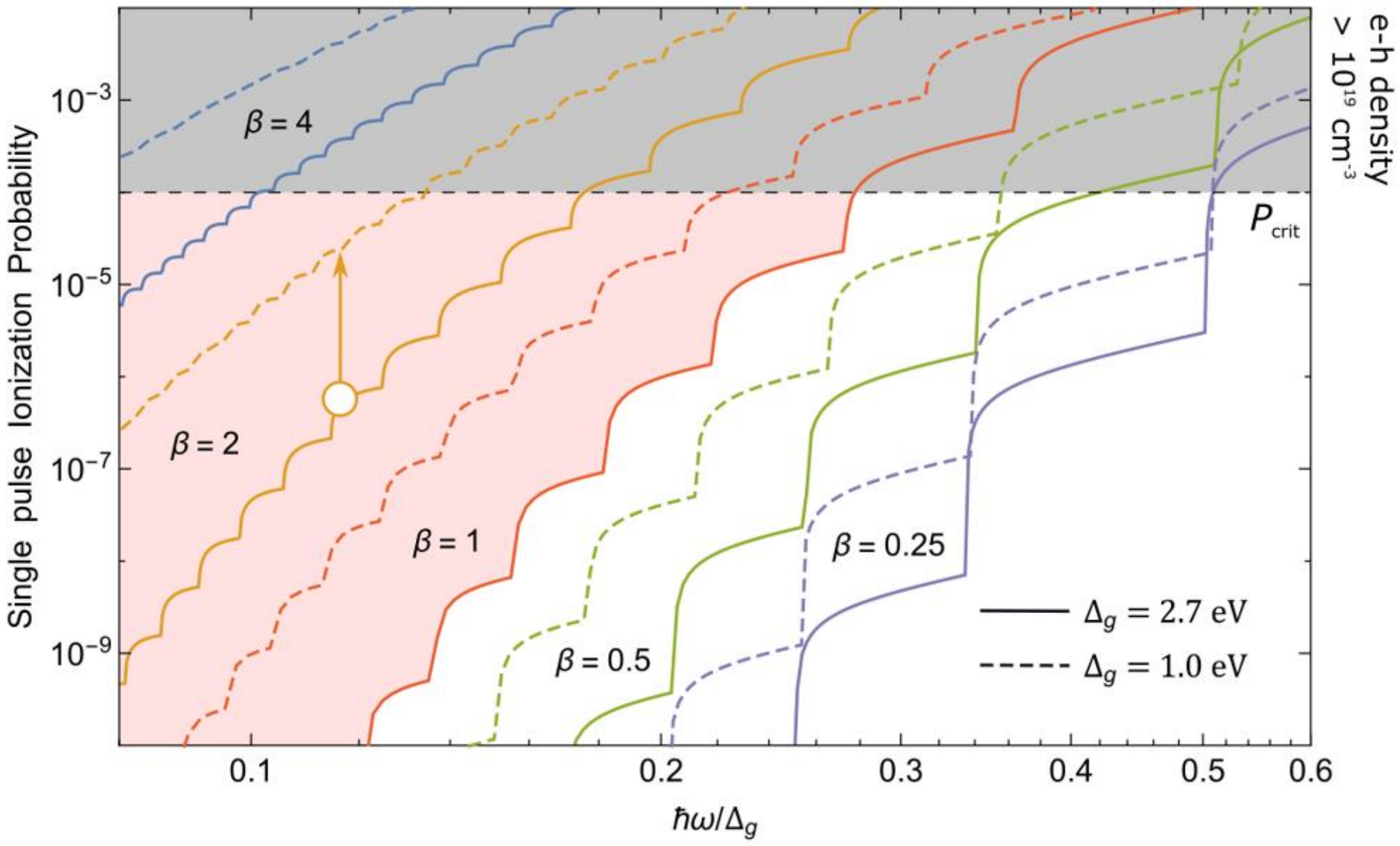

**Supplementary Fig. S1 | Single-pulse ionization probability estimates.** A system with ionization potential $\Delta_g = 2.7\ eV$ is excited by $t = 100$ fs pulse with a photon energy $\hbar\omega$ and peak electric field $E$ for various values of $\beta = 4, 2, 1, 0.5, 0.25$ (blue, orange, red, green and purple solid curves, respectively). The circle denotes the estimated experimental conditions, the shaded pink area denotes the region of strong-field ($\beta > 1$) excitation and the gray area denotes the dense photon-induced e-h plasma regime ($\rho > 10^{19}\ cm^{-3}$). The dashed lines show the same family of curves for $\Delta_g = 1.0\ eV$.

## 2. Sample fabrication

Crystalline GaP layer (~400 nm) is first grown on a GaAs substrate with an AlGaInP buffer layer by metal-organic chemical vapor deposition (MOCVD). Then this structure is directly bonded to a sapphire substrate (150 μm) after depositing ~2 μm $SiO_2$ layers on top of both surfaces. The AlGaInP/GaAs substrate is then removed by wet etching. The fabrication of the GaP nanostructures starts with a standard wafer cleaning procedure (using acetone, isopropyl alcohol and deionized water in that sequence under sonication), followed by $O_2$ and hexamethyl disilizane (HMDS) priming in order to increase the adhesion between GaP and subsequent spin-coated electron-beam lithography (EBL) resist of hydrogen silsesquioxane (HSQ). After spin-coating of HSQ layer with a thickness of ~200 nm, EBL and development in 25% tetra-methyl ammonium hydroxide (TMAH) were carried out to define the patterns in the HSQ resist. Finally, inductively-coupled plasma reactive ion etching (ICP-RIE) with $N_2$ and $Cl_2$ gases was used to transfer the HSQ patterns to the GaP layer and generate the GaP nanostructures; see Fig. 1b for a scanning electron micrographs (SEM) of the sample's fragment. The orientation of the GaP crystal lattice with respect to the metasurface is visualized by orienting the [001] direction perpendicular to the plane of metasurface, and then tilting the normal to the metasurface' plane by 15° toward the [111] direction of the GaP crystal lattice.

## 3. High harmonic measurement setup

In Supplementary Fig. S2, a detailed schematic of the optical setup used for high harmonic generation is shown.  The Extreme Mid-IR (EMIR) optical parametric amplifier (OPA) is a homebuilt $KNbO_3$/KTA 3-crystal/3-pass OPA. EMIR is pumped by The Ohio State University's GRAY laser, a homebuilt 80-fs Ti:Sapphire chirped pulse amplification system with a central wavelength of 780 nm and 4 mJ per pulse. The repetition rate of EMIR can be varied nearly continuously between 1 and 500 Hz using an external Pockels-cell-based pulse picker. EMIR can generate 200-fs mid-IR pulses with up to 25 μJ per pulse. The output wavelength of EMIR can be varied continuously from $\lambda = 2.7\ \mu$m to 4.5 $\mu$m. For the experiments, the MIR (idler) beam was fixed at $\lambda = 3.95\ \mu$m and collimated to a size of about 2.5 mm. Output modes were characterized for several different wavelengths using a WinCamD-FIR2-16-HR 2 to 16μm Beam Profiler System. The MIR pulse duration was measured using an AGS-crystal-based MIR autocorrelator for 3 and 3.6 μm to be $\tau = 200$ fs. The MIR spectra were obtained using an A.P.E. Wavescan USB MIR spectrometer. The MIR pulse energy was controlled with a half-wave plate—wire-grid polarizer pair in the range of about 1 to 6 μJ.

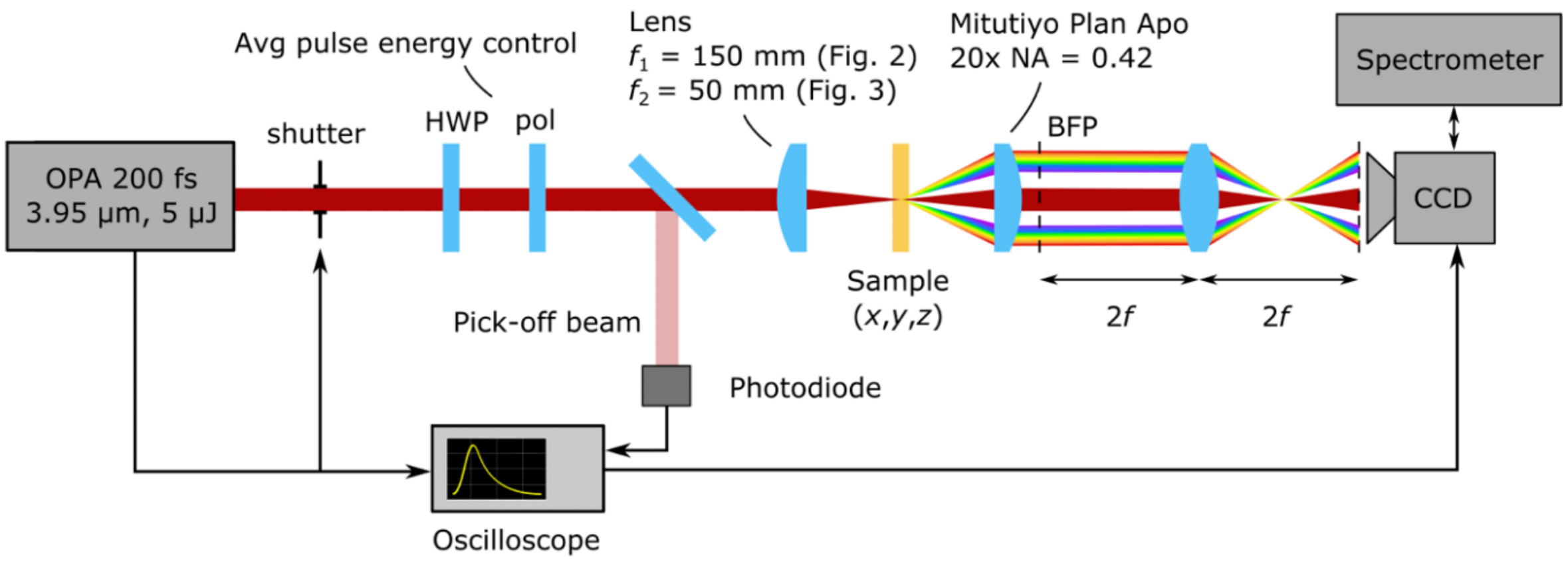

**Supplementary Fig. S2 | High harmonic generation setup.** OPA — optical parametric amplifier, HWP — half-wave plate, pol — wire-grid polarizer, BFP — back focal plane of the collecting objective.

## 4. Beam parameters

The beam spot at the sample plane was characterized using the same beam profiler system. For multi-pulse (perturbative) measurements, a $CaF_2$ lens with a focal distance of $f_{\mathrm{MP}}^{(1)} = 150$ mm was used; the focal spot size was measured to be $\Delta x_{\mathrm{FWHM}}^{(1)} = 175$ μm by $\Delta y_{\mathrm{FWHM}}^{(1)} = 153$ μm (intensity full width at half-maximum). For single-pulse (non-perturbative) measurements, a lens with a focal spot of $f_{\mathrm{SP}}^{(2)} = 50$ mm was used; the focal spot size was measured to be $\Delta x_{\mathrm{FWHM}}^{(2)} = 53$ μm by $\Delta y_{\mathrm{FWHM}}^{(2)} = 43$ μm (intensity full width at half-maximum). The peak intensity $I_0$ within the focal spot can be estimated approximating the beam profile with a 2D Gaussian:

$$W = I \int \exp\left(-\frac{4\ \mathrm{ln}2\ x^2}{\Delta x_{\mathrm{FWHM}}^2} - \frac{4\ \mathrm{ln}2\ y^2}{\Delta y_{\mathrm{FWHM}}^2}\right) dxdy,$$

Where $W$ is the pulse power. The energy-to-peak-intensity conversion coefficient is therefore expressed as:

$$K = \frac{I}{E} = \frac{4\ \mathrm{ln}2}{\pi \Delta x_{\mathrm{FWHM}} \Delta y_{\mathrm{FWHM}} \tau},$$

with the estimated values of $K_{\mathrm{SP}}^{(1)} = 1.65 \times 10^{16}\ \mathrm{s^{-1}cm^{-2}}$ for multi-pulse measurements and $K_{\mathrm{MP}}^{(2)} = 1.94 \times 10^{17}\ \mathrm{s^{-1}cm^{-2}}$ for single pulse-measurements. This way, the full range of accessed intensities ranging from 40 GW/cm$^2$ to 80 GW/cm$^2$ in the multi-pulse measurements and from 200 GW/cm$^2$ to 600 GW/cm$^2$ in the single-pulse measurements.

## 5. HHG measurements

Upon transmission through the sample, the upconverted signal was collected with a large-working distance (20 mm) Mitutoyo objective (NA = 0.42) which allows collection the transmitted harmonics as well as several diffraction orders. The back focal plane of the objective was projected onto the sensor of a thermoelectrically cooled back-illuminated CCD camera (Princeton Instruments PIXIS 1024 BUV). To spectrally filter the individual optical harmonics for back focal plane imaging, a set of long-, short- and band-pass filters (Thorlabs FEL and FESH series, FGB37) was used. Supplementary Fig. S3 shows the back focal plane images of harmonics H4, H5, H6, H7, H9 after spectral filtration, exposing the diffraction patterns, as well as the incoherent luminescence that fills the whole aperture of the objective for some wavelength ranges. Exposure times were 3 s, 100 ms, 10 s, 3 s, 10 s, respectively.

For HHG spectra acquisition, the back focal plane of the objective lens was projected onto the entrance slit of a monochromator (Chromex 250SM scanning monochromator) coupled to the same CCD camera. In a typical raw spectroscopic image, see Supplementary Fig. S4, the left image shows the camera output in the spectral range capturing H6 and H7, where both can be discerned on top of the luminescence background. In order to subtract the incoherent background, for each wavelength, we fit the $y$-section of the image to a Gaussian near the zeroth order diffraction (middle 30 pixels). For Fig. 2(b) of the main text, the amplitude of the Gaussian is plotted to separate the HHG signal from the luminescence background as a function of the wavelength.

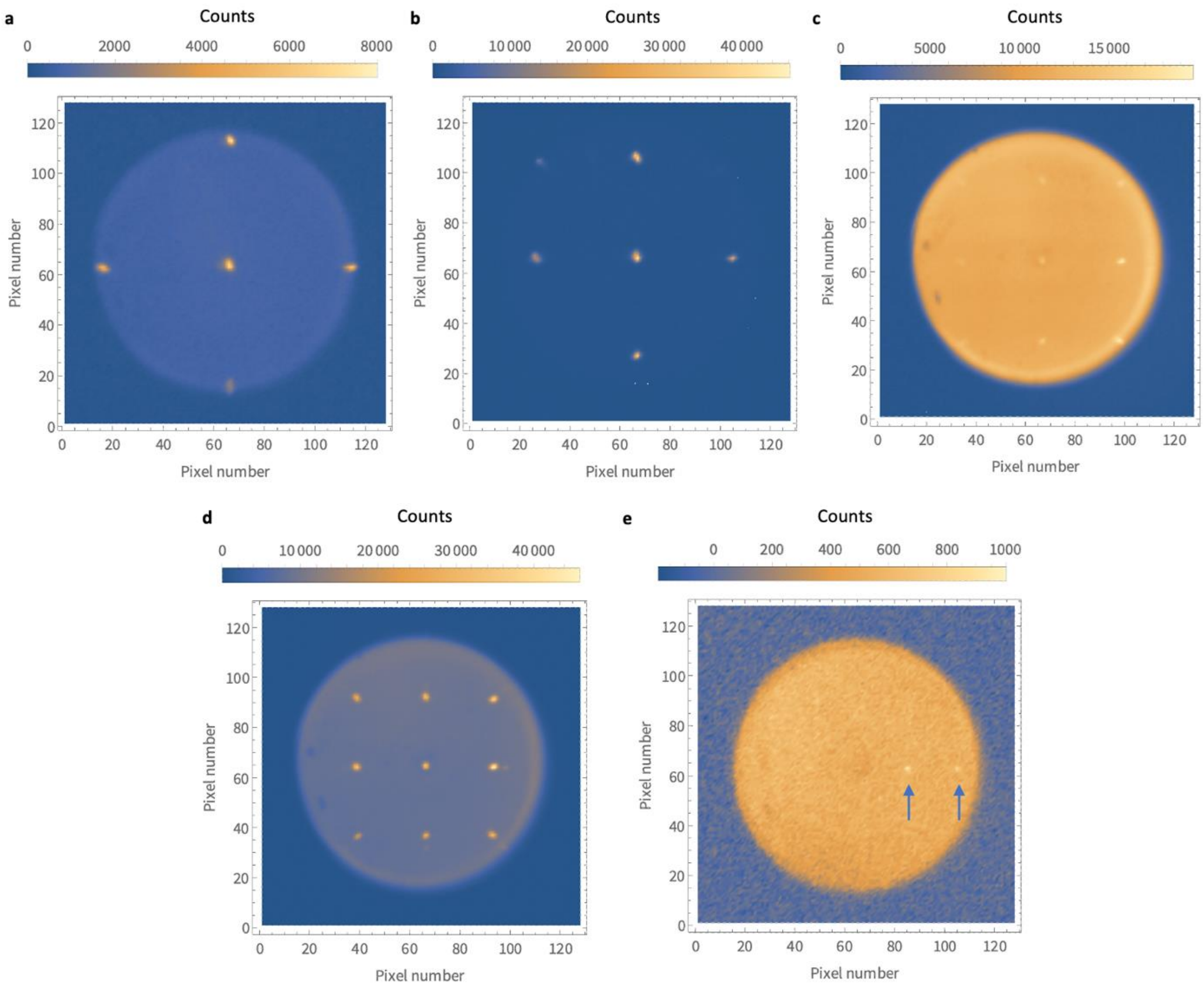

**Supplementary Fig. S3 | Back focal plane images of harmonics within different spectral bands. a,** $\lambda_{pass} > 900$ nm ($\lambda_{H4} \approx 990$ nm). **b,** $700\ nm < \lambda_{pass} < 850\ nm$ ($\lambda_{H5} \approx 790$ nm). **c,** $600\ nm < \lambda_{pass} < 700\ nm$ ($\lambda_{H6} \approx 660$ nm). **d,** $500\ nm < \lambda_{pass} < 700\ nm$ ($\lambda_{H7} \approx 560$ nm), **e,** $350\ nm < \lambda_{pass} < 500\ nm$ ($\lambda_{H9} \approx 430$ nm); arrows indicate the visible H9 diffraction orders. In some of the images, the luminescence background is present, filling the whole back focal plane of the objective.

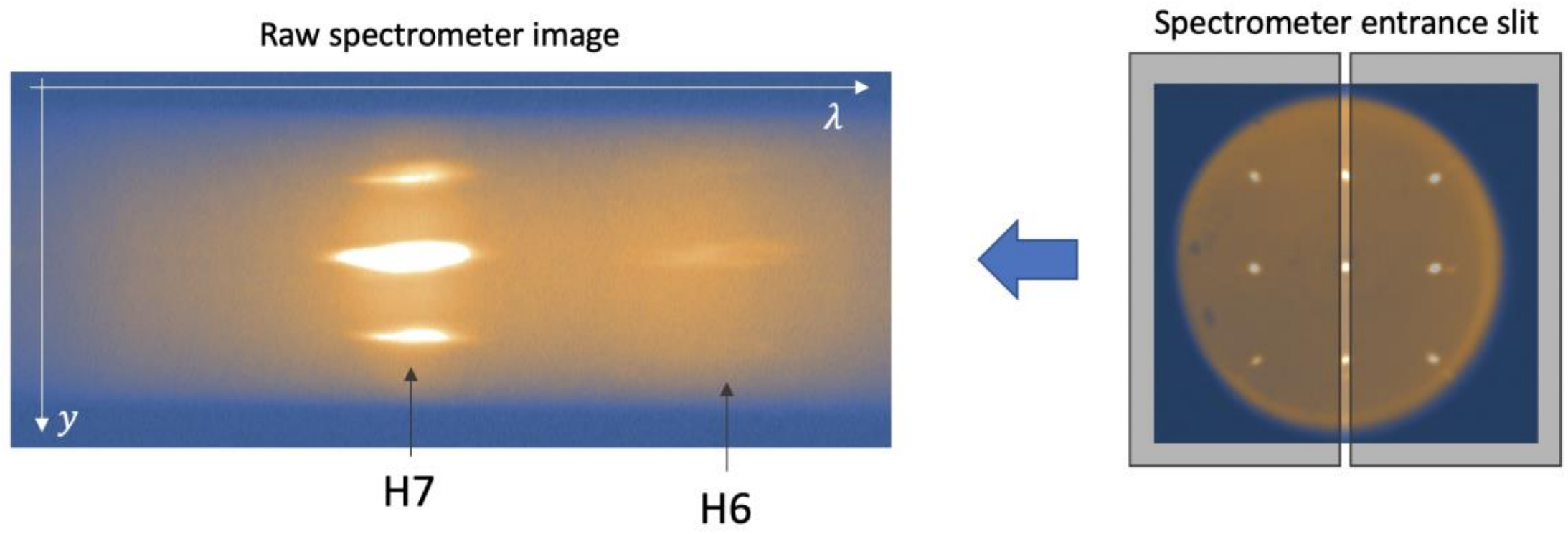

**Supplementary Fig. S4 | HHG spectroscopy schematic.** The back focal plane image of the sample was projected onto the entrance slit of the monochromator (right) resulting in the two-dimensional image at the camera after the monochromator (left). On the raw image, H7 and H6 are visible, both their (0; 0) and $(0; \pm 1)$ orders, as well as the broadband luminescence spectrum.

## 6. Single-shot HHG and damage threshold measurements

To reach the non-perturbative intensities with our experiments, we added a functionality for the setup to be able to irradiate the sample with individual laser pulses the emitted harmonics generation at substantially higher intensities than those used in multi-pulse measurements. The focusing lens was changed to one with $f = 50$ mm, and the beam size was measured to be an ellipse with axes $\Delta x_{\mathrm{FWHM}} = 53$ µm and $\Delta y_{\mathrm{FWHM}} = 43$ µm by putting a 2D micro-bolometer array sensor (DataRay WinCamD-IR-BB, pixel pitch 17 µm) in the focal point of an attenuated beam. For the single-pulse acquisition, the repetition rate of the laser was lowered to 10 Hz, and the measurements were done in the back focal plane setting with triggered exposure. The software-controlled trigger from the laser was sent to the mechanical shutter (1/30 s opening time) and an oscilloscope that received the signal from an amplified PbS photodiode that detected the energy of a pick-off pulse. The diode was calibrated using a pyroelectric power meter (Gentek-EO QE-B), averaging over 5000 pulses for each power setting in the range from 0.5 μJ to 5 μJ. Since the fluences used in these experiments lie close to the single-pulse damage threshold of the sample, we chose a fresh spot of the metasurface for each shot, moving at least 50 μm away between the shots. As an example, Supplementary Fig. S5 shows the resonant metasurface that has been exposed by single pulses in the energy ranges from 0.5 μJ to 3.8 μJ. We have divided the outcomes of single-pulse irradiation into three scenarios: state '0' with no apparent damage done to the sample, state '1' with the HSQ mask getting detached (as supported by the SEM images) and state '2' with the GaP resonators partially removed from the substrate. Supplementary Fig S5 shows the dependence of the outcome on the measured single-pulse energy. The importance of the single-pulse measurements is pinpointed by the fact that under multiple pulse irradiation, the sample gets severely damaged even at the periphery of the beam, where the intensity is very low: note the large crater at the bottom of Supplementary Fig S5, where the shutter was accidentally opened for several seconds, allowing about 30-50 pulses through at a moderate average pulse energy of about 2 μJ.

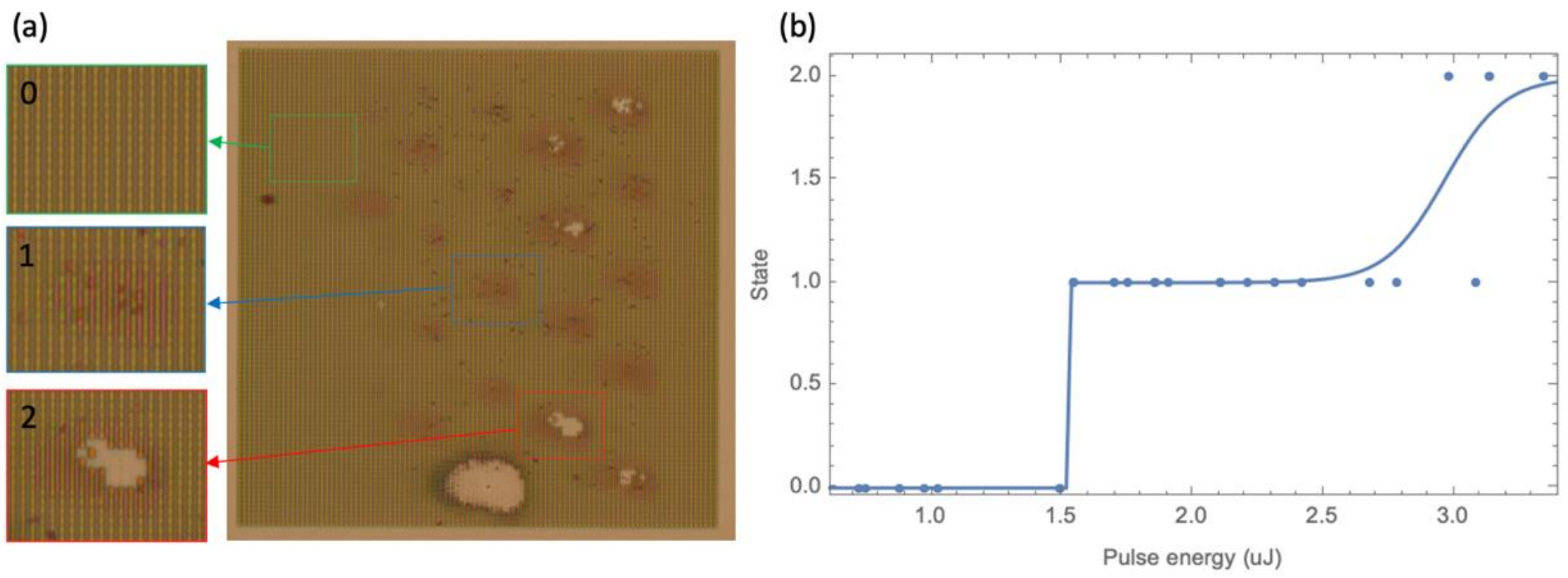

**Supplementary Fig. S5 | Cases of single-pulse damage in GaP metasurfaces. a,** An *a posteriori* optical image of sample #5 (resonant case), showing the different scenarios of pulse-metasurface interaction: from no damage at low fluences (scenario '0') to HSQ mask damage (scenario '1') to structural damage (scenario '2'). In '1,' the HSQ mask is detached from the surface of the sample, leaving the GaP resonators intact; the detached HSQ mask patches can be seen scattered around the sample as black dots. Scenario '2' describes a partial removal of the resonators in the center of the beam, with the bare substrate visible underneath. **b,** Damage threshold measurements. The dependence of the scenario (final state) number on the pulse energy shows two transitions characterized by the onset of mask detachment (around 1.5 μJ) and partial resonator ablation (around 2.5 μJ). The solid line is a double-logistic fit of the experimental data given with blue dots.

## 7. Conversion efficiency estimates

A source of cw radiation at $\lambda_{cw}^{(1)} = 532$ nm, with a measured power of $P_{\mathrm{cw}}^{(1)} = 80\ \mu$W right after the focusing lens, was attenuated by a stack of neutral density filters with the measured transmittance of 0.11 (OD1 filter), 0.01 (OD2 filter) and $7 \cdot 10^{-5}$ (OD4 filter), with the combined attenuation of $T^{(1)} = 7.7 \cdot 10^{-8}$, yielding the overall power at the sample site of $P_{\mathrm{cw}}^{(1)} T^{(1)} = 6.2$ pW. This beam then passed through the rest of the setup and was detected by the camera sensor. Under the exposure time of $t_{\mathrm{cw}}^{(1)} = 1$ s, the camera yielded $C_{\mathrm{cw}}^{(1)} = 1.4 \cdot 10^{6}$ counts at the zeroth diffraction order. Therefore, assuming the linearity of the signal over the exposure time, the sensitivity of the detection system can be estimated at $S^{(1)} = C_{\mathrm{cw}}^{(1)} / P_{\mathrm{cw}}^{(1)} T^{(1)} t_{\mathrm{cw}}^{(1)} = 2.3 \cdot 10^{17}$ cts/J. The H7 ($\lambda_{\mathrm{H7}} = 565$ nm) emitted from the sample area #5 yielded $C_{\mathrm{H7}} = 5.2 \cdot 10^{5}$ counts per $t_{\mathrm{H7}} = 1$ s exposure window, which includes only the zeroth diffraction orders; other orders have, both reflected and transmitted, not been taken into account in the efficiency calculations. The 7th harmonic power collected from the sample is therefore $P_{\mathrm{H7}} = \frac{C_{\mathrm{H7}}}{S^{(1)} t_{\mathrm{H7}}} \approx 2.3 \cdot 10^{-12}$ W. The average MIR power used in the experiment was equal to $P_{\mathrm{MIR}} = 1$ mW (2 μJ per pulse at a repetition rate of 500 Hz); the conversion efficiency of H7 is estimated calculated as $\eta_{\mathrm{H7}} = \frac{P_{\mathrm{H7}}}{P_{\mathrm{MIR}}} \approx 2.3 \cdot 10^{-9}$. From the relative intensities of the harmonics of different orders given in Fig. 2b, taking into account the spectral response of the detector (princetoninstruments.com/products/pixis-family/pixis) and objective (https://www.edmundoptics.com/p/20x-mitutoyo-plan-apo-infinity-corrected-long-wd-objective/6625/), we can estimate the following values for the conversion efficiencies of other harmonics: $\eta_{\mathrm{H4}} = 1.7 \cdot 10^{-10},\ \eta_{\mathrm{H5}} = 2.6 \cdot 10^{-8},\ \eta_{\mathrm{H6}} = 1.0 \cdot 10^{-11},\ \eta_{\mathrm{H9}} = 2.1 \cdot 10^{-12}$.

A similar procedure was carried out in the single-pulse case. Here, the calibrating laser was used at $\lambda_{\mathrm{cw}}^{(2)} = 633$ nm, close to the wavelength of the fifth harmonic ($\lambda_{\mathrm{H5}} = 790$ nm). The cw power measured in the focal plane of the focusing lens was $P_{\mathrm{cw}}^{(2)} = 166$ μW before attenuation. The attenuating filters in use were OD1 and OD6 with the measured combined transmittance of $T^{(2)} = 6 \cdot 10^{-8}$ and $P_{\mathrm{cw}}^{(2)} T^{(2)} = 10$ pW of cw power in the focal plane. A $t_{\mathrm{cw}}^{(2)} = 100$ ms exposure yielded $C_{\mathrm{cw}}^{(2)} = 2.5 \cdot 10^{5}$ counts at the camera. The detection system sensitivity can be estimated at $S^{(2)} = C_{\mathrm{cw}}^{(2)} / P_{\mathrm{cw}}^{(2)} T^{(2)} t_{\mathrm{cw}}^{(2)} = 2.5 \cdot 10^{17}\ \mathrm{cts/J}$. A single-shot exposure of the sample to a pulse with an energy of $E_{\mathrm{MIR}} =\ 1$ μJ yielded $C_{H5} = 2.6 \cdot 10^{5}$ counts of H5 signal, totaling $E_{\mathrm{H5}} = \frac{C_{\mathrm{H5}}}{S^{(2)} t_{\mathrm{H5}}} = 10^{-12}$ J of detected H5 energy. The conversion efficiency can therefore be estimated as $\eta_{5\omega} = \frac{E_{\mathrm{H5}}}{E_{\mathrm{MIR}}} = 10^{-12}\ \mathrm{J} / 10^{-6}\ \mathrm{J} = 10^{-6}$; adjusting for the spectral sensitivity of the camera and the objective at 633 nm and 790 nm results in $\eta_{5\omega}^{\mathrm{adj}} = 1.4 \cdot 10^{-6}$ which almost two orders of magnitude larger than that of the multi-pulse case. Supplementary Table S2 provides a comparison between the conversion efficiencies for the 5th and the 7th harmonics in various solid-state HHG systems.

| Material | Harmonic order | Conversion efficiency | Efficiency per 1 μm thickness |
|---|---|---|---|
| GaP metasurface [this work] | 5 (SP) | $1.4 \cdot 10^{-6}$ | $3.5 \cdot 10^{-6}$ |
| | 7 (MP) | $2 \cdot 10^{-9}$ | $5 \cdot 10^{-9}$ |
| ZnO [S3] | 5 | $3 \cdot 10^{-5}$ | $10^{-7}$ |
| | 7 | $6 \cdot 10^{-6}$ | $2 \cdot 10^{-8}$ |
| Periodically poled LiNbO3 [S4] | 5 | $10^{-2}$ | $4 \cdot 10^{-7}$ |
| | 7 | $10^{-2}$ | $4 \cdot 10^{-7}$ |
| Si metasurface [S5] | 5 | $5 \cdot 10^{-9}$ | $2.2 \cdot 10^{-8}$ |
| ENZ material [S6] | 5 | $10^{-8}$ | $1.3 \cdot 10^{-7}$ |
| | 7 | $10^{-10}$ | $1.3 \cdot 10^{-9}$ |

**Supplementary Table S2 | 5th and 7th harmonic conversion efficiencies in various previously reported solid-state HHG systems, compared to the efficiencies observed in the GaP metasurface.**

## 8. Tensor analysis of harmonics' properties

Owing to the stark difference in the nonlinear susceptibility tensor structures for the even- and odd-order harmonics, their emission efficiencies and polarization states can be strongly dependent on driver polarization and crystal orientation. Figure 2d of the main text shows the intensity of H7 and H6 as a function of analyzer angle for one of the diffracted orders, with the analyzer placed after the collection objective. While H7 is polarized along the pump radiation (90°/270° direction), H6 is elliptical, with the main semi-axis directed at around 45° with respect to the pump polarization. Also, Fig. 2b shows disproportionality between the even and odd harmonics' intensities. The difference in the polarization response and conversion efficiencies between the even- and odd-order harmonics can be qualitatively explained in terms of the nonlinear tensor symmetries. For the odd-order processes:

$$\chi^{(2k+1)}_{i_1 i_2 \dots i_{2k+1}} \neq 0;\ i_1 = i_2 = \dots = i_{2k+1},$$

meaning one can reasonably expect the main contribution from terms $I^{(2k+1)} \propto \chi^{(2k+1)}_{i_1 i_2 \dots i_{2k+1}}$.

However, due to the zincblende crystal structure of GaP, in the bulk for the even-order processes,

$$\chi^{(2k)}_{i_1 i_2 \dots i_{2k}} = 0;\ i_1 = i_2 = \dots = i_{2k},$$

meaning the main contributions to even-order harmonics will come from multiple off-diagonal components, opening potential to various polarization states of the output harmonics. We have analyzed the symmetry-enabled components of zincblende crystal nonlinear susceptibility tensors up to the 6th order; see Supplementary File 1. In the configuration where the input field is in the form of $E = \frac{E_0}{\sqrt{2}}(1, 1, 0)$ in the frame of the crystal structure, the nonlinear polarization for odd-order processes is in the form of $P_{\text{odd}} = \frac{P_{\text{odd},0}}{\sqrt{2}}(1, 1, 0)$, where $E_0$ and $P_{\text{odd},0}$ are constants. In contrast, for even-order processes, the polarization is in the form of $P_{\text{even}} = P_{\text{even},0}(0, 0, 1)$, where $P_{\text{even},0}$ is a constant. This means that if the GaP (100) plane were to lie in the plane of the metasurface, the nonlinear polarization at even harmonics would have been oriented out of metasurfaces' plane and weakly coupled to the zeroth diffraction order. Because of the 15° tilt of the normal to the metasurfaces' plane toward the [111] direction of GaP, some of this polarization outcouples to the zeroth order and can be detected. Nevertheless, the efficiency of this process is not optimized, and this fact serves to explain the low relative efficiency of the even harmonics in our experiment. Both the even- and odd-harmonic efficiency can be enhanced by judiciously choosing the crystal structure orientation. In Supplementary Fig. S6, the absolute values of the fifth and sixth order nonlinear polarizations, as well as their projections on the GaP crystal structure frame (see panel **a** for designations), are plotted as a function of the tilt angle $\theta$ ($\theta = 15^{\circ}$ in the experiment). Here, for the sake of simplicity, we assumed that for all nonzero tensor components, $\chi^{(5)}_{ijklmn} = \chi^{(5)}_{0}$ and $\chi^{(6)}_{ijklmno} = \chi^{(6)}_{0}$. Panels **b**, **c** show the nonlinear polarization components generated in the metasurfaces along the $x$, $y$ and $z$ for H6 and H5, respectively. Changing $\theta$ from $15^{\circ}$ to about $90^{\circ}$ can boost the even harmonic output by almost two orders of magnitude, as well as the odd harmonic output by a factor of 17.

The polarization state differences of the even and odd harmonics can be qualitatively understood as well. For the odd-order processes, at $\theta \approx 0^{\circ}$ the major contributors to the emitted harmonics' polarization state are $P_x$ and $P_y$ components, where $P_x \approx P_y$, generating harmonic beams with approximately the same polarization state as the pump beam. In the even harmonic case, at $\theta \approx 0^{\circ}$, $P_x \approx P_y \approx 0$, and $P_z$ becomes the primary polarization. As $\theta$ increases, the contributions from all three directions may become similar in amplitude, generating polarization states that are elliptical and may not necessarily be aligned with the pump polarization, as seen in Fig. 2d of the main text. The interplay between the relative orientations of the crystal lattice and the metasurface lattice is a promising topic of the future studies.

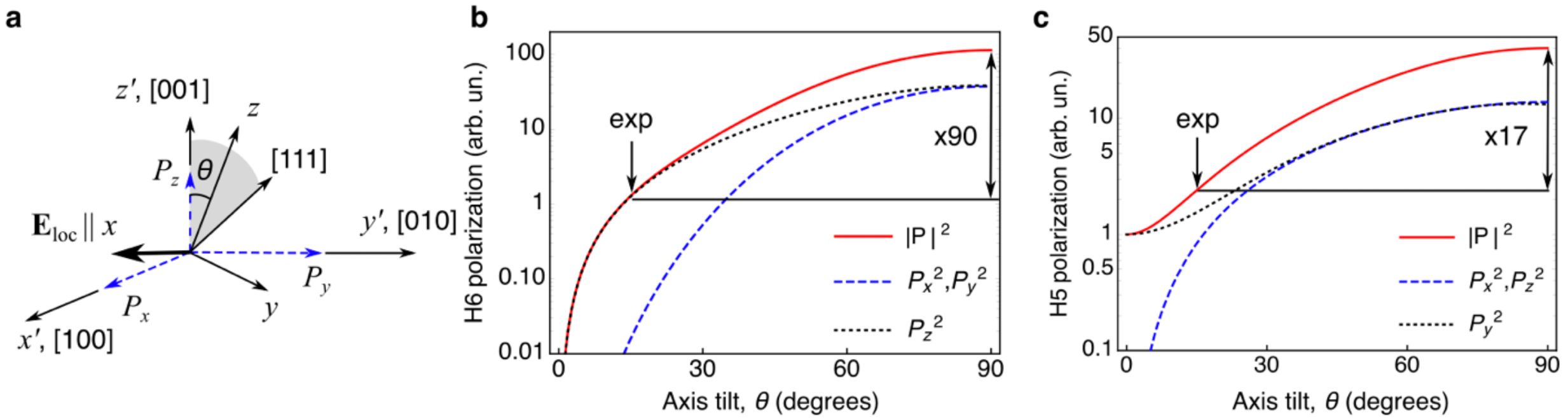

**Supplementary Fig. S6 | Optimizing the HHG output by a judicious choice of the crystal lattice orientation in GaP metasurfaces. a,** A model of local fields within the metasurface $\boldsymbol{E}_{loc}$ in the GaP crystal structure frame $(x', y', z')$. The direction of the average local fields within the metasurface coincides with the $x$ direction in Fig. 1 of the main text. **b**, **c**, Polarizations of H6 and H5, respectively, as functions of the tilt angle $\theta$ between the normal to the metasurfaces' plane $(z)$ and [111] direction of the GaP lattice. Solid red lines correspond to the total polarizations; the dashed blue lines correspond to polarizations along $x'$ and $y'$ (in **b**) and along $x'$ and $z'$ (in **c**); the dotted black lines correspond to polarizations along $z'$ (in **b**) and $y'$ (in **c**). The experiments were conducted for $\theta = 15°$. An overall enhancement of the nonlinear polarization by a factor of 90 for the 6th harmonic and by a factor of 17 for the 5th harmonic is observed if $\theta = 90°$.

Supplementary references:

[S1] Perelomov, A. M., Popov, V. S., Terent'ev, M. V. Ionization of atoms in an alternating electric field. *Sov. Phys. JETP* **23**, 924 (1966).

[S2] Keldysh, L. V. Ionization in the field of a strong electromagnetic wave. *Sov. Phys. JETP* **20**, 1307 (1965).

[S3] Gholam-Mirzaei, S. *et al.* High harmonic generation in ZnO with ahigh-power mid-IR OPA. *Appl. Phys. Lett.* **110**, 061101 (2017).

[S4] Hickstein, D. D. *et al.* High-harmonic generation in periodically poled waveguides. *Optica* **4**, 1538 (2017).

[S5] Liu, H. *et al.* Enhanced high-harmonic generation from an all-dielectric metasurface. *Nat. Phys.* **14**, 1006–1010 (2018).

[S6] Yang, Y. *et al.* High-harmonic generation from an epsilon-near-zero material. *Nat. Phys.* **15**, 1022–1026 (2019).